\documentclass[12pt]{revtex4}
\usepackage{color,graphicx,url}
\usepackage{times,amsmath,amssymb,amsfonts,amsthm,subfigure}
\usepackage[colorlinks, urlcolor=black, citecolor=black, linkcolor=black]{hyperref}

\date{\today}
\begin{document}

\title{Difficulties in applying pure Kohn-Sham density functional
  theory electronic structure methods to protein molecules}
  \author{Elias~Rudberg}
  \email{elias.rudberg@it.uu.se}
  \affiliation{\mbox{Division of Scientific Computing,}
    \mbox{Department of Information Technology,}
    \mbox{Uppsala University,}
    Box 337,
    SE-751 05 Uppsala,
    Sweden}

\begin{abstract}
Self-consistency based Kohn-Sham density functional theory (KS-DFT)
electronic structure calculations with Gaussian basis sets are
reported for a set of 17 protein-like molecules with geometries
obtained from the protein data bank. It is found that in many cases
such calculations do not converge due to vanishing HOMO-LUMO gaps.
A sequence of polyproline I helix molecules is also studied, and it is
found that self-consistency calculations using pure functionals fail
to converge for helices longer than six proline units. Since the
computed gap is strongly correlated to the fraction of Hartree-Fock
exchange, test calculations using both pure and hybrid density
functionals are reported. The tested methods include the pure
functionals BLYP, PBE, and LDA, as well as Hartree-Fock and the hybrid
functionals BHandHLYP, B3LYP, and PBE0.  The effect of including
solvent molecules in the calculations is studied, and it is found that
the inclusion of explicit solvent molecules around the protein
fragment in many cases gives a larger gap, but that convergence
problems due to vanishing gaps still occur in calculations with pure
functionals. In order to achieve converged results, some modeling of
the charge distribution of solvent water molecules outside the
electronic structure calculation is needed. Representing solvent water
molecules by a simple point charge distribution is found to give
non-vanishing HOMO-LUMO gaps for the tested protein-like systems also
for pure functionals.
\end{abstract}

\maketitle

\section{Introduction}

Kohn-Sham density functional theory~\cite{hist-kohnsham} (KS-DFT) has
been widely used in electronic structure calculations. Efficient
algorithms have been developed that allow KS-DFT methods to be applied
for large molecules, including hundreds and even thousands of
atoms~\cite{linsca-g}. However, a problem with KS-DFT is the
self-interaction error (SIE). For example, it is known that SIE causes
severe errors in computed polymer
polarizabilities~\cite{SIE-polymer-polarizabilities-2003} where the
problem becomes more and more severe with increasing system size.
Thus, application of KS-DFT to large
systems is not always straightforward.

One important application of KS-DFT for large molecules is the study
of proteins, whose properties are of interest in biology. In this
work, we study the applicability of standard 
self-consistency based
KS-DFT methods for
calculations on protein molecules.

\section{Method \label{sec:method}}

In KS-DFT methods, the electron density is expressed via a set of
orbitals in a similar way as in the Hartree-Fock~\cite{book-szabo}
(HF) method.  We consider here non-periodic spin-restricted KS-DFT methods at zero
electronic temperature. Then, the number of occupied orbitals is
$n_{occ} = n/2$ where $n$ is the number of electrons in the system. The
Kohn-Sham orbitals are determined by solving
\begin{equation} \label{eq:ks}
\mathbf{F} \mathbf{C} = \mathbf{S} \mathbf{C} \mathbf{\epsilon}
\end{equation}
where $\mathbf{F}$ is the Kohn-Sham matrix, $\mathbf{C}$ the matrix of
orbital coefficients, $\mathbf{S}$ the overlap matrix and
$\mathbf{\epsilon}$ the diagonal matrix of orbital energies. The
matrices in \eqref{eq:ks} are $N \times N$ matrices, where $N$ is the
number of basis functions. Given a set of $N$ orbitals that constitute
a solution to \eqref{eq:ks}, a set of occupied orbitals is formed by
including the $n_{occ}$ orbitals of lowest energy. The occupied
orbitals determine the density matrix $\mathbf{D}$ as
\begin{equation} \label{eq:densitymatrix}
D_{ij} = 2 \sum_{k=1}^{n_{occ}} C_{ik} C_{jk}
\end{equation}
where the columns of $\mathbf{C}$ are taken to be ordered by the
corresponding orbital energies. The Kohn-Sham matrix $\mathbf{F}$ is
computed from $\mathbf{D}$ according to the chosen
exchange-correlation functional. Since $\mathbf{F}$ depends on
$\mathbf{D}$, an iterative procedure is used to find a self-consistent
solution.

Calculations where a new density matrix is computed by occupying the
orbitals of lowest energy as described above are in this work referred
to as \emph{self-consistency based} calculations, to clearly
distinguish them from direct minimization approaches.

In self-consistency based calculations, convergence schemes such as
damping~\cite{damping} and DIIS~\cite{pulay82} are usually employed,
where a new Kohn-Sham matrix is constructed by taking information from
previous iterations into account. See the work of Kudin and Scuseria
\cite{sc-ks} for an overview of such convergence schemes.

The self-consistency based approach usually works well provided that there
is a sizable gap between the highest occupied molecular orbital (HOMO)
and lowest unoccupied molecular orbital (LUMO) energies. However, there is no
guarantee that the HOMO-LUMO gap will be large; the gap depends on the
studied system as well as on the basis set and on the used
exchange-correlation functional. If the gap is very small the
procedure of determining the occupied orbitals needed in~\eqref{eq:densitymatrix} becomes ill-defined and
then a self-consistent solution may be difficult or even impossible to
find.

KS-DFT exchange-correlation functionals can be divided into two main
classes: pure and hybrid functionals. In hybrid functionals, some
fraction of HF exchange is added to the Kohn-Sham matrix, often using
empirically determined constants.

\section{Results \label{sec:results}}

This section includes results of HF and KS-DFT calculations for
various protein-like systems. The presented results were computed
using the Ergo program~\cite{m-ergo}. The obtained results can however
be readily reproduced using other KS-DFT codes employing Gaussian
basis functions.

The convergence scheme used in the reported calculations is a
combination of damping and DIIS, as implemented in the Ergo
program. This scheme essentially uses damping in early iterations,
with a dynamically adapted damping factor such that the step size is
decreased whenever the energy goes up, leading to very small steps in
difficult cases. However, details of this scheme do not affect the
reported results; the observed convergence problems are due to
vanishing HOMO-LUMO gaps, a problem that neither damping nor DIIS-like
schemes can resolve.

In this work, calculations were considered ``converged'' when the
largest absolute matrix element of the matrix commutator
$\mathbf{F}\mathbf{D}\mathbf{S}-\mathbf{S}\mathbf{D}\mathbf{F}$ was
smaller than $5 \times 10^{-4}$. This particular choice of convergence
threshold is however not critical for the reported results, since the
calculations that failed to converge due to vanishing gaps were
typically very far from reaching this criterion.

\subsection{Molecules from the protein data bank \label{sec:pdbdirect}} 

Table~\ref{tbl:results_631gss} shows computed HOMO-LUMO gaps for a set
of protein-like molecular systems with geometries taken from the
protein data bank (PDB)~\cite{pdb}. 
In cases where the PDB file contains more than one structure, the one
labeled ``model 1'' was used.
The 17 structures in Table~\ref{tbl:results_631gss} were selected in
order to give examples of various types of protein-like systems, with
the requirement that positions of hydrogen atoms should be included in
the PDB file. The net charge of each molecule, shown in the fourth
column in Table~\ref{tbl:results_631gss}, was chosen after performing
a set of HF/3-21G calculations for different charges. For each system,
the charge that gave the largest HOMO-LUMO gap was chosen.
Calculations were performed using six
different KS-DFT functionals as well as HF. The employed density
functionals include the pure functionals LDA (SVWN5), BLYP, and PBE as
well as the hybrid functionals B3LYP, PBE0, and BHandHLYP with HF
exchange fractions of 20\%, 25\%, and 50\%, respectively. The Gaussian
basis set 6-31G** was used.

\begin{table} 
\begin{tabular}{l@{$\quad$}lrr@{$\qquad$}rrrrrrr}
\hline  
\hline
        &                      &        &        & \multicolumn{7}{c}{Computed HOMO-LUMO gap [eV]}    \\
\cline{5-11}
PDB ID  &  Type                &  atoms & charge & HF     & BHandHLYP &  PBE0 & B3LYP &  BLYP  &  PBE  &  LDA  \\
\hline
2P7R    & biosynthetic protein &   73   &   0    &  12.03 &      7.23 &  4.65 &  4.16 &   2.12 &  2.10 &  2.12 \\
1BFZ    & peptide              &   87   &   0    &  11.96 &      8.13 &  6.18 &  5.77 &   3.97 &  3.93 &  3.79 \\
2IGZ    & antibiotic           &  147   &   0    &  11.81 &      7.99 &  5.70 &  5.27 &   3.27 &  3.23 &  3.12 \\
1D1E    & neuropeptide         &  243   &  +3    &  10.14 &      6.05 &  3.96 &  3.47 &   1.56 &  1.58 &  1.54 \\
1SP7    & structural protein   &  352   &  +3    &   9.13 &      4.06 &  1.65 &  0.87 &        &       &       \\
1N9U    & signaling protein    &  182   &   0    &   9.12 &      3.80 &  1.12 &  0.57 &        &       &       \\
1MZI    & viral protein        &  225   &  -3    &   8.77 &      3.80 &  1.29 &  0.54 &        &       &       \\
1XT7    & antibiotic           &  217   &  +1    &   8.51 &      4.85 &  3.32 &  2.65 &   1.02 &  1.24 &  1.30 \\
1PLW    & neuropeptide         &   75   &   0    &   7.25 &      2.31 &  0.36 &  0.29 &        &       &       \\
1FUL    & peptide              &  135   &  -1    &   6.95 &      1.85 &  0.20 &  0.16 &        &       &       \\
1EDW    & peptide              &  399   &  -1    &   6.89 &      2.02 &  0.26 &  0.21 &        &       &       \\
1EVC    & bacterial toxin      &  109   &  -2    &   5.82 &      1.14 &  0.30 &  0.24 &        &       &       \\
1RVS    & de novo protein      &  172   &   0    &   5.60 &      0.60 &       &       &        &       &       \\
2FR9    & peptide toxin        &  194   &  -2    &   5.48 &      0.55 &  0.26 &  0.21 &        &       &       \\
2JSI    & hormone              &  198   &  -1    &   5.26 &      0.66 &  0.24 &  0.19 &        &       &       \\
1LVZ    & peptide-binding protein  &  185   &   0    &   5.05 &      0.71 &  0.31 &  0.25 &        &       &       \\
1FDF    & signaling protein    &  416   &  +1    &   3.64 &      0.25 &  0.13 &  0.11 &        &       &       \\
\hline  \hline
\end{tabular}
  \caption{Results of HF and KS-DFT calculations using the Ergo
    program on a set of protein-like molecules. Basis set:
    6-31G**. Blank space indicates that no converged result was
    obtained.
\label{tbl:results_631gss}}
\end{table}

The most important conclusion from the results shown in
Table~\ref{tbl:results_631gss} is that in many cases, calculations using pure functionals
fail to converge for molecules larger than a few hundred
atoms.
 Note that the blank spaces
in the columns for BLYP, PBE, and LDA indicate not only that no gap
value was obtained, but that those calculations did not give any
meaningful results at all since they failed to converge.

For the calculations that did converge, 
the computed HOMO-LUMO gaps are strongly correlated to the fraction of
HF exchange included in the functional, with a large fraction of HF
exchange giving a large gap. Thus, for each molecule, the HF method
yields the largest HOMO-LUMO gap, while the BHandHLYP functional
consistently gives a larger gap than PBE0 and B3LYP. The pure
functionals give much smaller gaps; in many cases, no converged
results were obtained for the pure functionals due to vanishing
HOMO-LUMO gaps. For the 1RVS system, the B3LYP and PBE0 calculations
also failed to converge.

To check the basis set dependence, calculations with larger basis sets
were also performed for the smaller systems. The larger basis set
results indicate that the computed HOMO-LUMO gaps are not critically
dependent on the basis set. For example, calculations using the
cc-pVTZ basis set for the 1BFZ, 1EVC, 1PLW, and 2P7R molecules gave
HOMO-LUMO gaps that differed by less than 15\% compared to the 6-31G**
results. In some cases, a larger basis set gives a smaller gap.

Calculations using the smaller basis set 3-21G were also
performed. Those results indicate that already the 3-21G basis set
gives similar gaps and convergence behavior for the different
functionals.

\subsection{Size dependence \label{sec:sizedependence}}

The results in Table~\ref{tbl:results_631gss} indicate that the
convergence problems due to small gaps are to a large extent system
dependent. However, for even smaller protein-like fragments,
consisting of only a few amino acids, calculations typically converge
without problems also for pure functionals. Therefore there is reason
to believe that that the convergence problems increase with increasing
molecular size.

To further assess the size dependence, calculations were also
performed for a sequence of polyproline I helix molecules of
increasing length. The model helix geometries were generated using the
Gabedit program~\cite{allouche_gabeditgraphical_2011}, applying the
``Build Polypeptide'' function with the ``Polyproline I'' conformation
followed by the ``add hydrogens'' command.

\begin{figure}
 \begin{center}
 \includegraphics[width=0.49\textwidth]{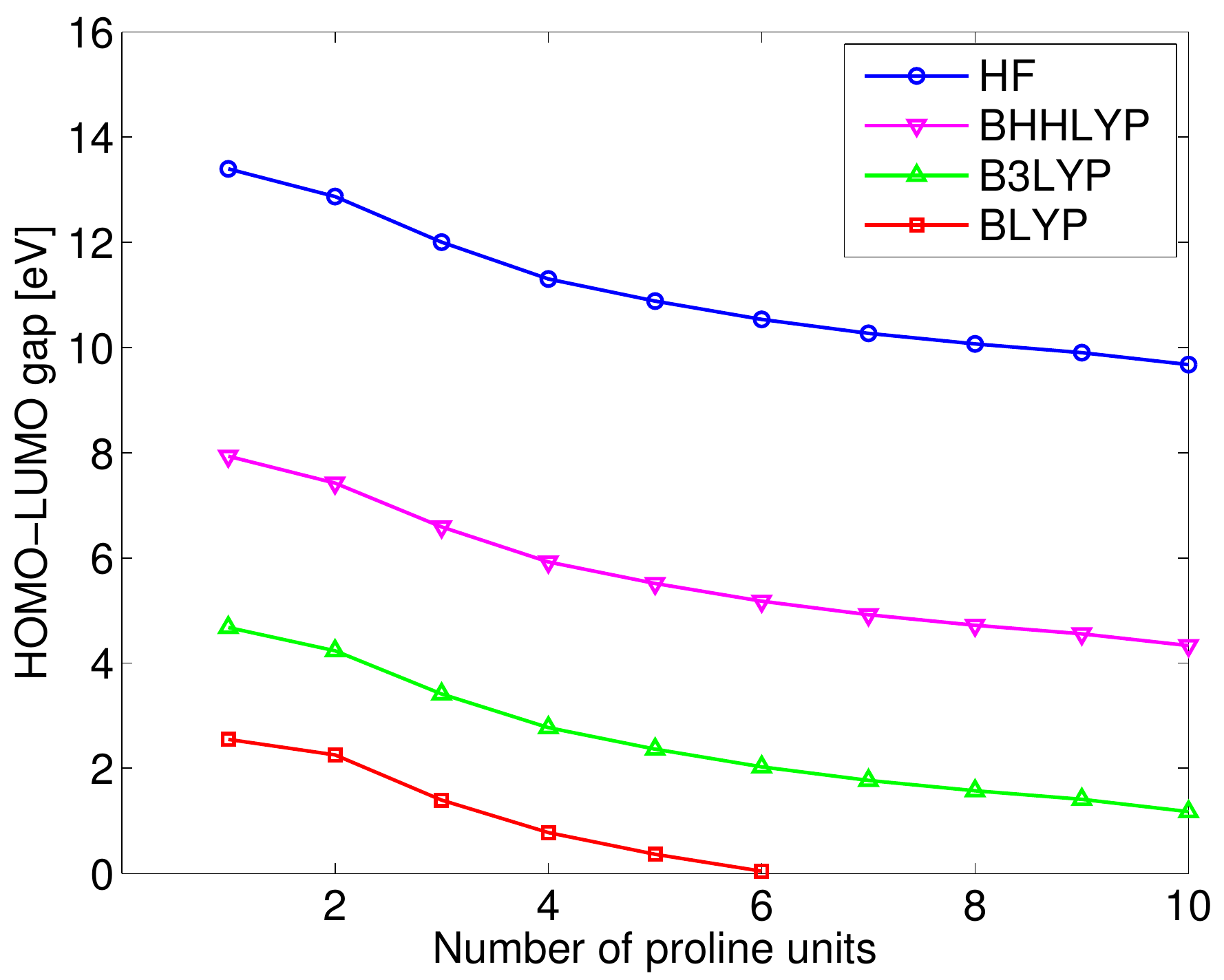}
 \end{center}
\caption{Computed HOMO-LUMO gaps for polyproline I helix
  molecules. Basis set: 6-31G**. The BLYP calculations for
  helices with 7-10 proline units failed to converge.
\label{fig:polyproline_i_gaps}
}
\end{figure}

Computed HOMO-LUMO gaps for the polyproline I helix systems obtained
using the KS-DFT functionals BLYP, B3LYP, and BHandHLYP as well as HF
are shown in Figure~\ref{fig:polyproline_i_gaps}. The size dependence
is clearly seen: for any given functional, the computed HOMO-LUMO gap
decreases with increasing helix length, and because the computed gaps
for pure functionals are so small, those calculations fail to converge
for sizes larger than six proline units.

As seen in Figure~\ref{fig:polyproline_i_gaps}, the problem of
vanishing HOMO-LUMO gaps is in this case clearly related to the system
size. However, the system size is not the only important factor. For
example, performing the corresponding test calculations for helices in
the polyproline II conformation gives sizable gaps even for very large
systems. Apparently the problem of vanishing gaps is not seen for the
fairly stretched out polyproline II helices, but the problem does
appear for the more compact polyproline I conformation.

\subsection{Including solvent water molecules \label{sec:includingsolvent}}

The calculations in sections~\ref{sec:pdbdirect}
and~\ref{sec:sizedependence} were done for isolated protein-like
systems without any surrounding water molecules. This is not
completely realistic, since in real biological systems protein
molecules are typically dissolved in water, and the solvent water
molecules can have a significant effect on both the molecular geometry
and the electronic structure of the protein. In this section, we
consider the effect of explicitly including solvent water molecules
when performing KS-DFT calculations for protein-like systems.

Since structures from the PDB in general do not include solvent
molecules, model structures including solvent molecules were generated
by molecular dynamics (MD) simulations at standard temperature and
pressure using the Gromacs program~\cite{gromacs-jctc-2008}. The
AMBER03 force field and the TIP3P water model were used. The MD
simulations were done with the ``position restraints'' option in the
Gromacs program, thus keeping the protein geometry reasonably close to
the original geometry from the PDB, but allowing some motion and
complete freedom of the surrounding solvent water molecules.

MD simulations were done for four of the systems from
Section~\ref{sec:pdbdirect}: 1FUL, 1LVZ, 1PLW, and 1RVS. For each of
them, a number of MD runs were performed, generating ten uncorrelated
MD snapshots. From each snapshot a model system with solvent was
created by including all water molecules within 4~{\AA} from the
solute. For comparison, corresponding model structures without solvent
were also generated for the same set of MD snapshots. The structures
without solvent differ slightly from the original PDB structures
as the molecules moved during the MD simulations.

\begin{figure}
 \begin{center}
 \subfigure[$\, $ Without surrounding water molecules \label{fig:four_pdb_mols_gaps_without_h2o}]{
 \includegraphics[width=0.49\textwidth]{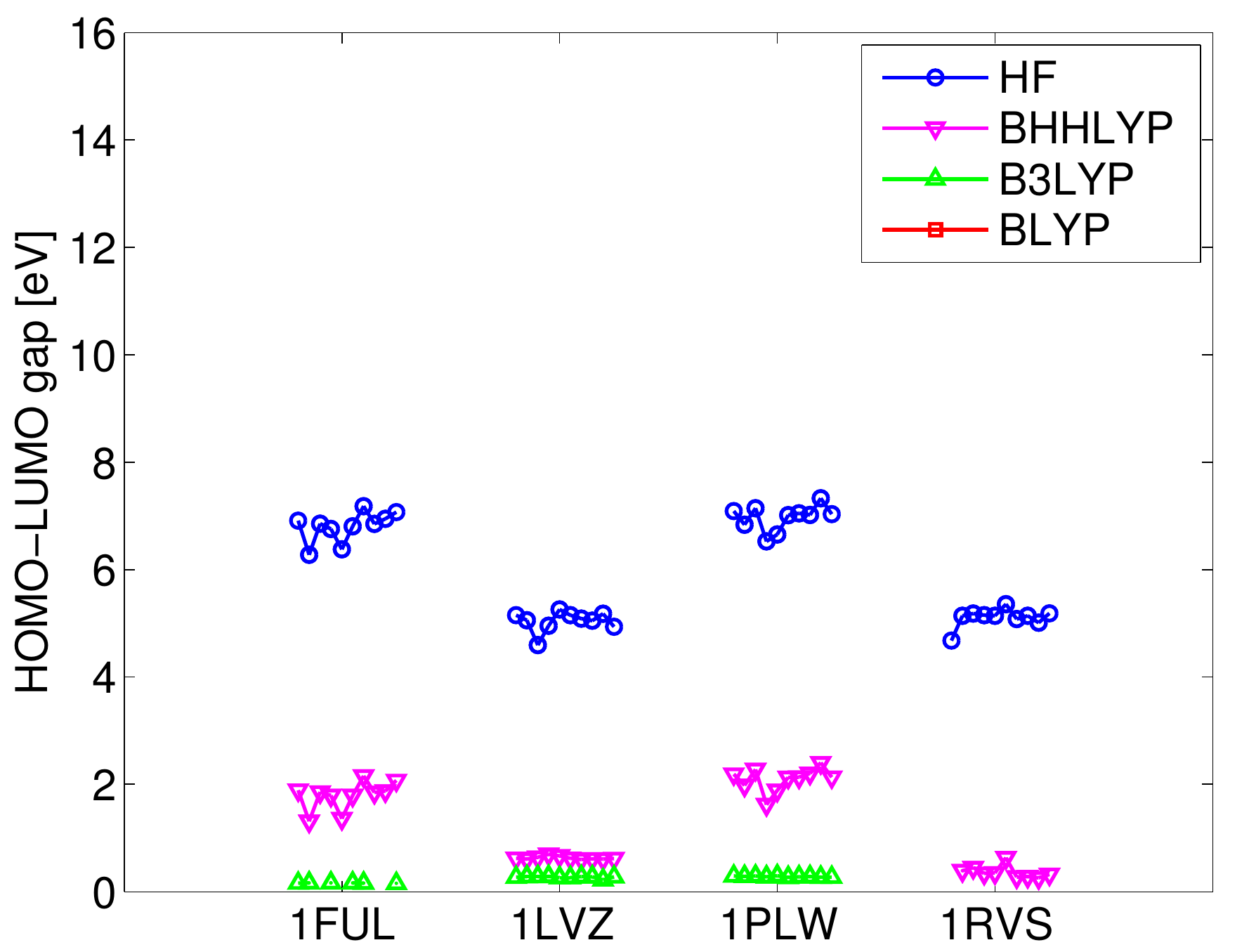}}
 \subfigure[$\, $ With surrounding water molecules \label{fig:four_pdb_mols_gaps_with_h2o}]{
 \includegraphics[width=0.49\textwidth]{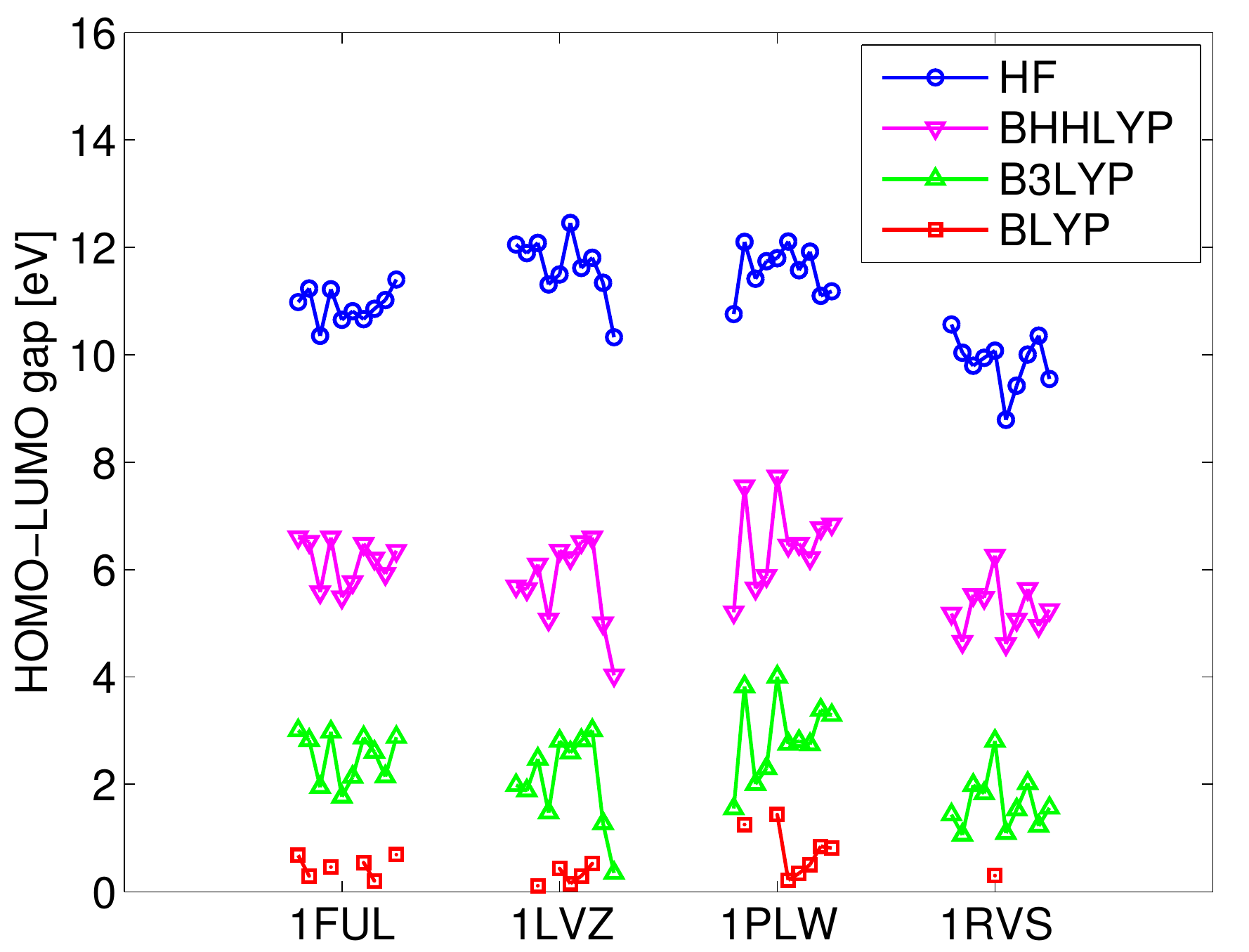}}
 \end{center}
\caption{Computed HOMO-LUMO gaps for protein-like systems without and
  with surrounding water molecules. Basis set: 3-21G. Several of the
  BLYP calculations failed to converge even with surrounding water
  molecules.
\label{fig:four_pdb_mols_gaps_with_and_without_h2o}
}
\end{figure}

Figure~\ref{fig:four_pdb_mols_gaps_with_and_without_h2o} shows
computed HOMO-LUMO gaps for the model systems generated from MD
simulations. To reduce the computational effort, these calculations
were done using the 3-21G basis set. Comparisons to larger basis set calculations
done for a few cases indicate that the effect of this limited basis
set is not critical; qualitatively similar results would probably be
obtained with a larger basis set.

Figure~\ref{fig:four_pdb_mols_gaps_without_h2o} shows computed gaps
for structures without surrounding solvent molecules. As can be
expected from the results of Section~\ref{sec:pdbdirect}, the BLYP
calculations here give vanishing gaps and therefore fail to
converge.

Figure~\ref{fig:four_pdb_mols_gaps_with_h2o} shows that the inclusion
of explicit solvent molecules in the calculation in general gives a
larger gap. However, in several cases the BLYP calculations still fail
to converge. There is some randomness; BLYP calculations may or may
not converge depending on the positions of included solvent molecules
in that particular MD snapshot.

\begin{figure}
 \begin{center}
 \includegraphics[width=0.49\textwidth]{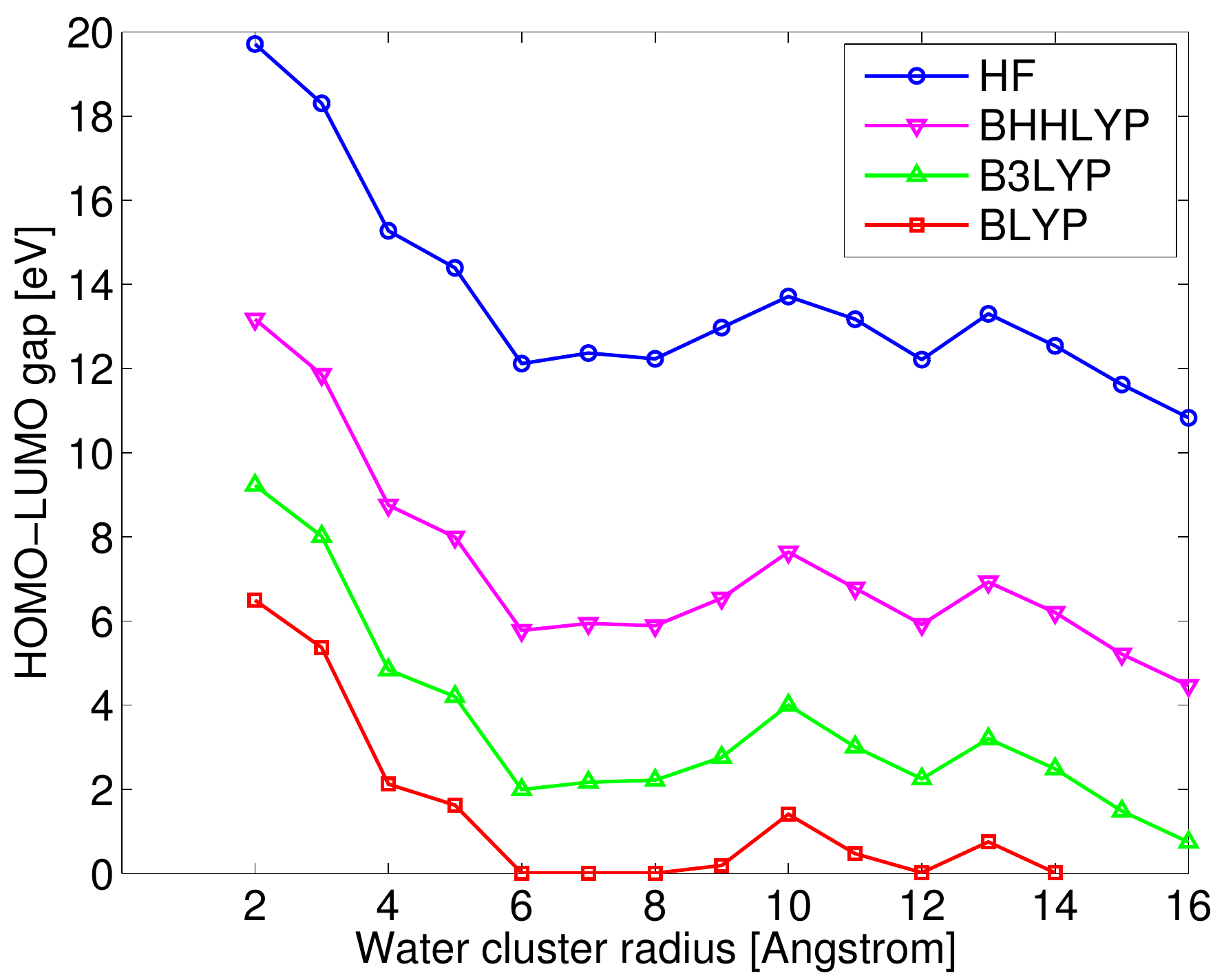}
 \end{center}
\caption{Computed HOMO-LUMO gaps for water clusters. Basis set: 3-21G. The BLYP calculations for water clusters of radius 15-16~{\AA} failed to converge.
\label{fig:h2o_clusters_gaps}
}
\end{figure}

In the test calculations presented in
Figure~\ref{fig:four_pdb_mols_gaps_with_h2o}, water molecules up to
4~{\AA} from the solute were included. One may of course include more
solvent molecules, but doing so does not seem to solve the problem. In
fact, vanishing gaps for pure functionals is a problem also when
considering water clusters, as shown in
Figure~\ref{fig:h2o_clusters_gaps}. The water cluster geometries were
generated by including all water molecules within a certain radius
from a snapshot from an MD simulation at standard temperature and
pressure. 
The problem of pure functionals giving vanishing gaps for water clusters was reported previously \cite{sparsity2011}.

The computed gaps in Figure~\ref{fig:h2o_clusters_gaps} 
decrease rather drastically at 13-16~{\AA} radius, but this is a coincidence
for the particular MD snapshot considered here; if continuing to
larger clusters using HF or hybrid functionals the gaps tend to
stabilize \cite{linmemHF,linmemDFT}. However, pure functionals
are not straightforwardly applicable for water clusters
generated in this way. Therefore, embedding a protein-like molecule in
water by including explicit water molecules up to some radius cannot
be expected to solve the convergence problems due to vanishing
gaps. In order to achieve converged results with pure functionals,
some modeling needs to be done of the other water molecules, outside
the domain of the electronic structure calculation, as will be seen in
the next section.

\subsection{Including point charges representing solvent water molecules outside computational domain \label{sec:withpointcharges}}

Previous work by Cabral do Couto et
al.~\cite{CabraldoCouto-gaps-brazilian-2004} has shown that for water
clusters extracted from a larger simulation, orbital energies are
strongly affected by the water molecules surrounding the clusters, and
that such surface effects can to some extent be corrected for by
including point charges representing the surrounding molecules. Cabral
do Couto et al. found that HOMO-LUMO gaps are significantly increased
when adding point charges representing surrounding water molecules. In
this section, the approach of adding such point charges is applied to
the case of protein molecules embedded in water.

The test systems used in this section are the same as those in
Section~\ref{sec:includingsolvent} except that now water molecules
outside the electronic structure calculation domain are included via
point charges. These ``outer'' water molecules are not explicitly
included in the electronic structure calculation, but they are
represented by point charges corresponding to their simple point
charge (SPC) distribution. That is, oxygen and hydrogen atoms
are represented by point charges of -0.82 and +0.41,
respectively. Outer water molecules up to 10~{\AA} away from the
studied system were included. This gives a large number of point
charges (for 1RVS, around 4800 point charges were used) but the
extra computational effort is anyway small since the point charges
only affect the core Hamiltonian matrix. The expensive Coulomb, HF
exchange, and exchange-correlation parts of the calculation are not
affected by the added point charges.

\begin{figure}
 \begin{center}
 \subfigure[$\, $ Without surrounding water molecules \label{fig:four_pdb_mols_gaps_without_h2o_with_spc}]{
 \includegraphics[width=0.49\textwidth]{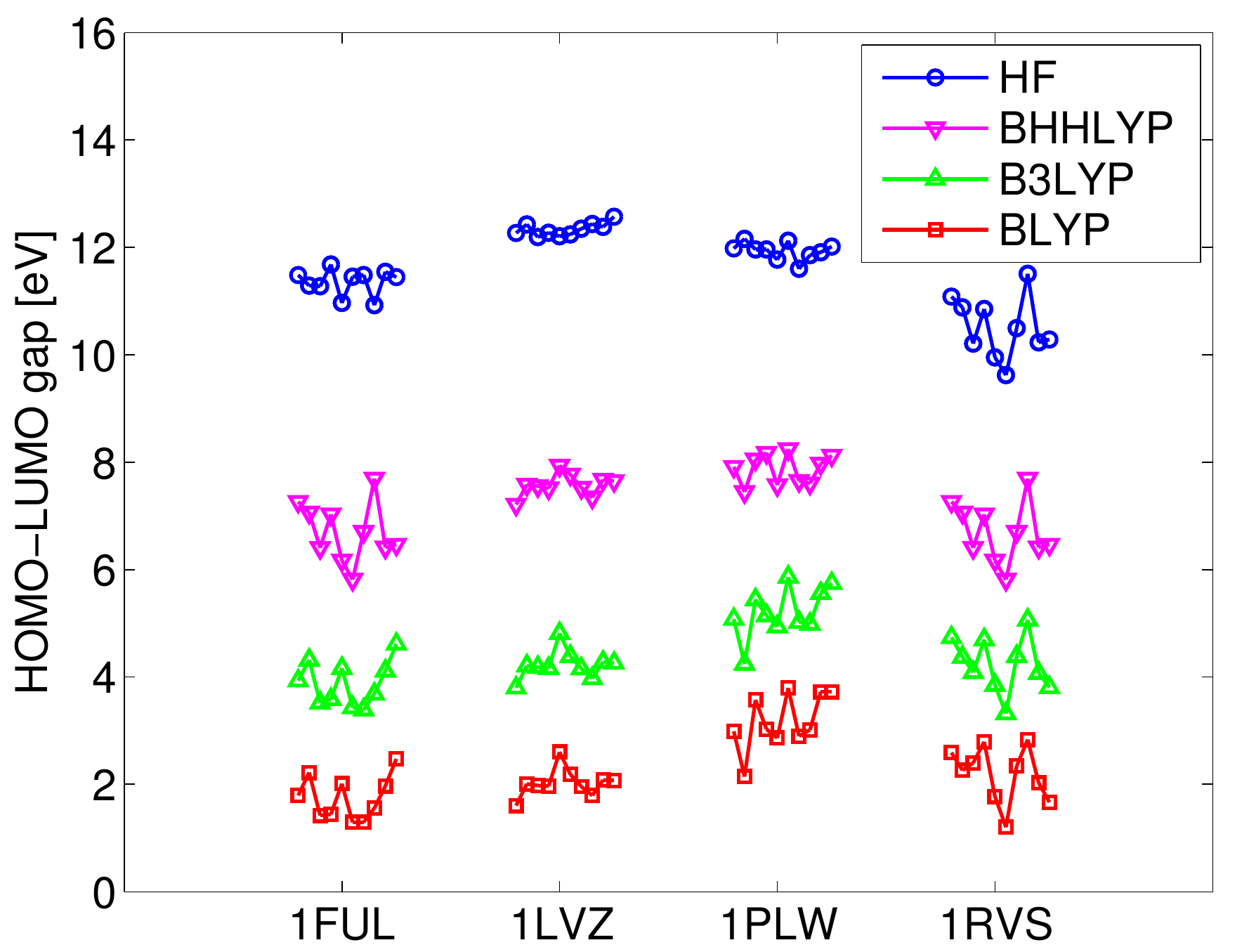}}
 \subfigure[$\, $ With surrounding water molecules \label{fig:four_pdb_mols_gaps_with_h2o_with_spc}]{
 \includegraphics[width=0.49\textwidth]{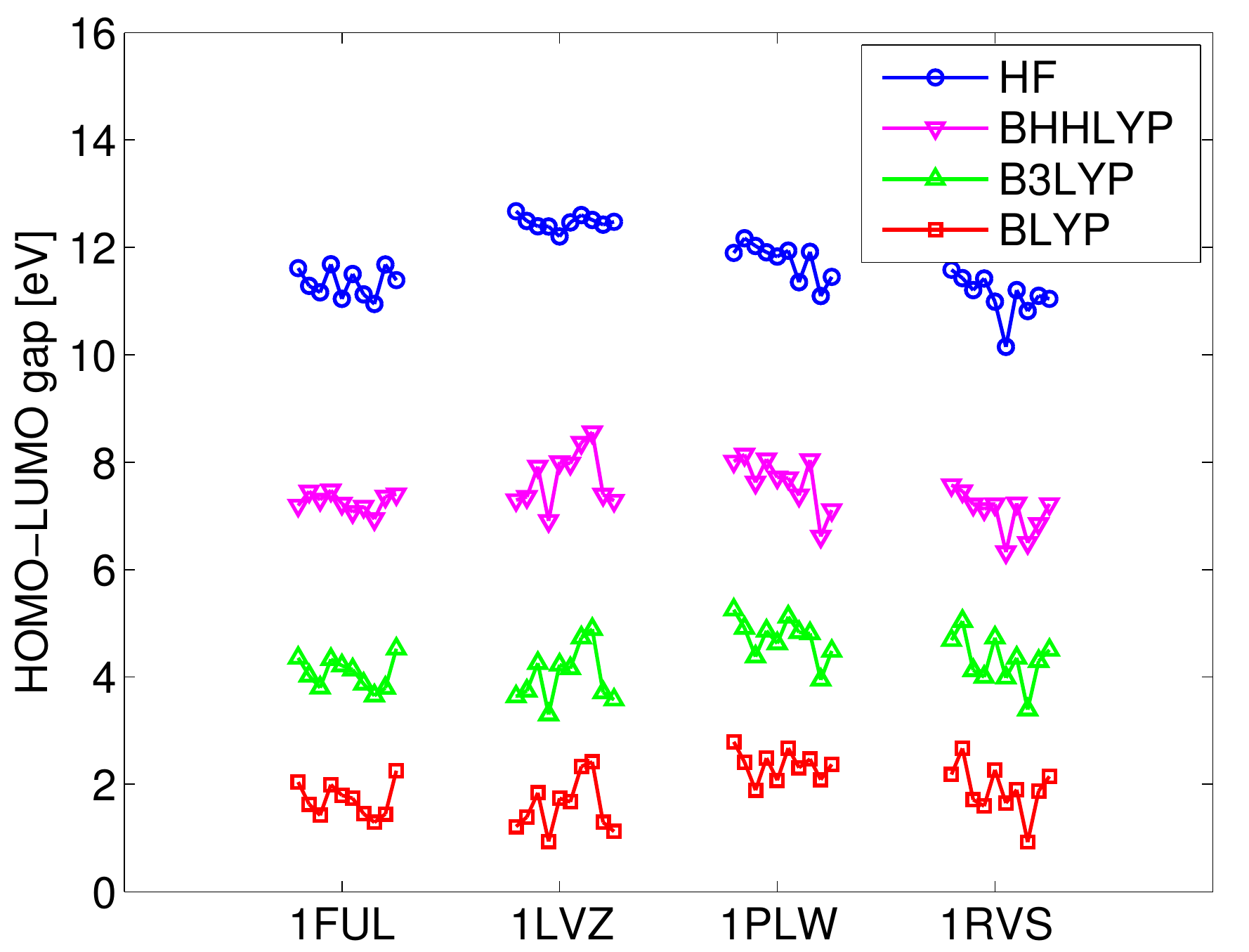}}
 \end{center}
\caption{Computed HOMO-LUMO gaps for protein-like systems without and
  with surrounding water molecules. Basis set: 3-21G. In both cases,
  water molecules outside the computational domain were represented by
  SPC point charges.
\label{fig:four_pdb_mols_gaps_with_and_without_h2o_with_spc}
}
\end{figure}

Figure~\ref{fig:four_pdb_mols_gaps_with_and_without_h2o_with_spc}
shows computed HOMO-LUMO gaps for the same systems as in
Figure~\ref{fig:four_pdb_mols_gaps_with_and_without_h2o}, but now
including SPC point charges as described above. Note that in the
calculations shown in
Figure~\ref{fig:four_pdb_mols_gaps_with_h2o_with_spc}, water molecules
are included in two ways: water molecules up to 4~{\AA} from the solute
are explicitly included in the electronic structure calculation, and
additional water molecules between 4 and 14~{\AA} away from the solute
are represented by point charges.

Judging from
Figure~\ref{fig:four_pdb_mols_gaps_with_and_without_h2o_with_spc}, the
approach of including point charges representing water molecules
outside the electronic structure calculation domain appears to solve
the convergence problems for pure functionals: when point charges are
included in this way, BLYP calculations give HOMO-LUMO gaps of more
than 0.9~eV in all studied cases. This approach also gives convergence
for the polyproline I helix systems considered
in Section~\ref{sec:sizedependence}.

Thus, it appears that despite the discouraging results of
sections~\ref{sec:pdbdirect} and~\ref{sec:sizedependence},
calculations using pure functionals can be done for protein-like
systems provided that surrounding solvent water molecules are
accounted for by somehow taking their charge distribution into
account. If solvent water molecules are explicitly included in the
electronic structure calculation, surface effects must anyway be
handled by including the charge distribution water molecules further
away.

In this section, surface effects were handled using point charges in
the same way as in the work of Cabral do Couto et
al \cite{CabraldoCouto-gaps-brazilian-2004}. This
is easily done from an implementation point of view and required only
a minor modification to the Ergo program~\cite{m-ergo} that was used
to perform the calculations. Another way of taking effects of the
surrounding water into account would be to use a polarizable continuum
model, although that possibility was not explored here.

The point charge embedding approach was here considered as a tool to
obtain a converged solution. Of course, such point charges are a very
crude approximation of the solvent atoms they are supposed to
represent. One should therefore be careful when interpreting results
of such calculations, in particular regarding the electronic
structure near the boundary where point charges were added.

\section{Concluding remarks}

All calculations reported in this work were
performed using Gaussian basis sets, far from the basis set limit.
To better assess the basis set dependence, it would be desirable to
also perform calculations with other types of basis sets, e.g. plane
waves.

The results obtained here for protein fragments are in line with
previous findings that pure KS-DFT functionals give vanishing gaps for
large polypeptide and water cluster systems~\cite{linmemDFT}. Also,
vanishing PBE gaps have been reported for plane-wave calculations on
semiconductors~\cite{PhysRevB.81.153203.PBEnogap}.

Although the problem of pure KS-DFT functionals underestimating
HOMO-LUMO gaps is well known in the
literature~\cite{PhysRevLett.51.1884.perdew.gapProblem,PhysRevB.37.10159.sham.gapProblem,PhysRevB.53.3764.levy.gapProblem,salzner-gap-problem-1997,PhysRevB.78.235104.gapProblem,PhysRevLett.105.266802.gapProblem},
to the author's best knowledge the resulting convergence problems in
self-consistency based calculations for protein-like molecules has received
little, if any, attention previously.

It should be noted that the calculations reported in this work were
done for finite model systems. That is, periodic boundary conditions
were not used. When using a finite model system to describe a protein
in water solution, the domain must be truncated somewhere, and it is
then important to handle surface effects in some way, for example as
described in Section~\ref{sec:withpointcharges}. In a periodic
calculation there is no boundary and thus no surface effects to worry
about. Periodic calculations using pure KS-DFT for proteins have been
reported for example by Sulpizi et
al.~\cite{Sulpizi-large-protein-dft-2007}.

The calculations in this work were all performed using the self-consistency approach, as described in
Section~\ref{sec:method}. Therefore, a non-vanishing HOMO-LUMO gap was
here necessary to achieve convergence. It should be noted that other
optimization schemes for KS-DFT calculations exist, where a
parametrization is used that ensures that the density matrix stays
idempotent (and has the correct number of electrons), but where there
is no guarantee that the orbitals defining the density are the ones
having the lowest orbital energies. Then, a converged solution could in
principle be found even if the gap vanishes. However, such approaches
were not used in the present work.

In the calculations reported in this work, significant effort was made
to reduce the risk that the reported results are dependent on any
particular choice of starting guess density. For cases that turned out
to be difficult to converge, repeated calculations using several
different starting guesses were tried, including densities obtained
with other functionals and other basis sets. In those cases where
``convergence failure'' is reported, this does not mean only that one
particular calculation failed to converge, but that all calculation
attempts using various starting guesses failed.

All results reported in Section~III are from spin-restricted (closed
shell) calculations. Additional spin-unrestricted calculations with
different alpha- and beta-spin densities as starting guesses were
performed for many of the studied cases. In those cases,
spin-unrestricted calculations did not resolve the convergence
problems.

Test calculations using level shifting~\cite{levelshift} were
performed for a few of the difficult cases in
sections~\ref{sec:pdbdirect} and~\ref{sec:sizedependence}. If
employing a large enough shift, a converged result can sometimes be
obtained. However, if the resulting density is used as a starting
guess for a calculation without any level shift, different orbitals
are occupied and convergence is not obtained. 
Also, the calculations
with level shifting are very sensitive to the starting guess. In cases
where the usual self-consistency based approach (without level shifting) fails due
to vanishing gaps, 
calculations employing level shifting may converge 
to any of many possible
final results with small differences in energy, depending on the
starting guess. Such solutions found using level shifting do typically
not obey the \emph{aufbau} principle; that is, the occupied orbitals are not
the ones having the lowest orbital energies.
This suggests that proper \emph{aufbau} solutions to the standard
Kohn-Sham model may not exist for these cases; compare for example to
the case of chromium carbide considered by Kudin et al
\cite{kud-scus-cances-scf-2002}.
In any case, using level shifting does
not seem to be a satisfactory solution to the convergence problems,
since the final result then becomes heavily dependent on the starting
guess.

Another way to achieve convergence in difficult cases
would be to employ 
fractional finite-temperature occupation numbers in
the same way as in calculations for
metals~\cite{PhysRevLett.79.1337.metals,PhysRevB.79.241103.metals}.
Alternatively, instead of standard KS-DFT methods one may 
consider employing 
the \emph{extended} Kohn-Sham model~\cite{cances-extended-ks-2001}, or using
GW theory~\cite{PhysRevLett.96.226402.GWtheory}.
However, application of such methods goes beyond the scope of the present
work.

Application of self-consistency based pure KS-DFT methods to protein-like molecules without
including solvent often leads to convergence problems due to vanishing
HOMO-LUMO gaps. Although such problems can be alleviated by including
solvent molecules, they indicate that the applicability of such pure
KS-DFT methods may be limited: if a protein-like system surrounded by
air or vacuum is to be studied, it is unclear to what extent self-consistency based pure
KS-DFT methods can be applied. Further investigation of this issue
remains a subject of future work.

\section*{Acknowledgements}

The author wishes to thank E.~R.~Davidson, E.~H.~Rubensson, and
P.~Sa{\l}ek for helpful discussions.
The computations were performed on resources provided by the Swedish
National Infrastructure for Computing (SNIC) at High Performance
Computing Center North (HPC2N) and Uppsala Multidisciplinary Center
for Advanced Computational Science (UPPMAX).

\bibliography{biblio} \bibliographystyle{apsrev} 

\end{document}